# Development of an automatic modification system for generated programs using ChatGPT

[1]Jun Yoshida    [2]Oh Sato    [3]Hane Kondo    [4]Hiroaki Hashiura    [5]Atsuo Hazeyama

[1, 2, 3, 5]Tokyo Gakugei University, 4-1-1 Nukuikita-machi, Koganei-shi, Tokyo 184-8501 JAPAN

[4]Nippon Institute of Technology, 4-1 Gakuendai, Miyashiro-machi, Minamisaitama-gun, Saitama 345-8501 JAPAN

E-mail: [1] e205415f@st.u-gakugei.ac.jp, [2] m228112p@st.u-gakugei.ac.jp, [3] m228110m@st.u-gakugei.ac.jp, [4] hashiura@nit.ac.jp, [5] hazeyama@u-gakugei.ac.jp

**Abstract** In recent years, the field of artificial intelligence has been rapidly developing. Among them, OpenAI's ChatGPT excels at natural language processing tasks and can also generate source code. However, the generated code often has problems with consistency and program rules. Therefore, in this research, we developed a system that tests the code generated by ChatGPT, automatically corrects it if it is inappropriate, and presents the appropriate code to the user. This study aims to address the challenge of reducing the manual effort required for the human feedback and modification process for generated code. When we ran the system, we were able to automatically modify the code as intended.

**Keywords** ChatGPT, Programming, System development, Automatic modification

## 1. Introduction

In recent years, the development of artificial intelligence (AI) has been progressing rapidly. Among these advancements, OpenAI's ChatGPT [1], based on the GPT-4 (Generative Pre-trained Transformer 4) architecture [2], excels in natural language processing tasks and can generate source code, theoretically enabling the creation of web applications. However, the generated code often has issues related to consistency and program rules. In such cases, it is necessary to identify and correct each error in the generated output, which then needs to be regenerated. In this study, we developed a system that tests the code generated by ChatGPT, automatically corrects errors if it is inappropriate, and presents the appropriate code to the user. This system aims to reduce the manual effort required for human feedback and the correction process for generated code.

## 2. Related Work

In the context of utilizing ChatGPT for programming and system development, we discuss related research.

### 2.1. Research by White et al. [3]

White et al. presented methods to solve common issues in a pattern format when automating general engineering activities using large language models (LLMs) like ChatGPT. While their study focused on requirements extraction and system design, it did not specifically address code auto-correction.

### 2.2. Research by Fill et al. [4]

In experiments conducted by Fill et al., tasks were defined and presented in the form of Entity Relationship (ER) diagrams and JSON examples. Their study focused on the upstream process of software development by having ChatGPT interpret ER diagrams in JSON format. However, it did not address downstream processes like coding or correction of code.

### 2.3. Research by Fukuda et al. [5]

Fukuda et al. proposed an automatic review method for software design documents using large language models. Their study focused on reviewing existing documents but did not address the entire process from code generation to testing and regeneration.

### 2.4. Positioning of This Study

This study focuses on coding using ChatGPT and highlights the novelty of automatically generating test code, running tests, and correcting code based on the results, unlike the related studies.

## 3. Proposed System

### 3.1. Mechanism

The proposed system leverages ChatGPT's natural language understanding and code generation capabilities. Typically, users copy and paste the code generated by ChatGPT into their environment, check for compilation and runtime errors, and send the errors back to ChatGPT for corrections. This system automates this process by using test code generated alongside the main code to execute and catch errors, sending them back to ChatGPT for corrections until the code runs without errors.

## 3.2. System Overview

The system operates as follows:
1. The user inputs the content for the code to be generated.
2. The system requests ChatGPT via its API.
3. ChatGPT generates the code and corresponding test code.
4. The system compiles the generated code and runs tests.
5. If errors are found, the system sends the code and errors back to ChatGPT.
6. Steps 2-5 are repeated until no errors are found.
7. The final code is presented to the user, with limits on iteration and execution time to prevent resource exhaustion.

Figure 1 shows an overview of the system. This system is intended for programmers and is not recommended for learners.

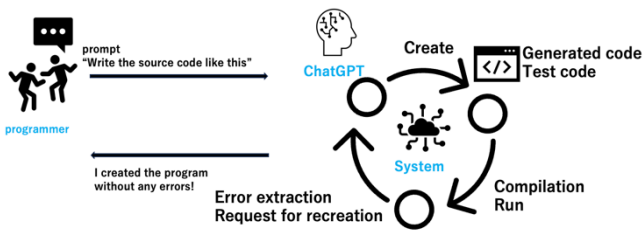

**Figure 1　System overview**

## 3.3. System Requirements

The system must meet the following requirements:

### 3.3.1. Main Functions
- The system can accept prompts from users.
- The system can connect to the ChatGPT API.
- The system can accurately send requests to the ChatGPT API.
- The system can accurately receive responses from the ChatGPT API.
- The system can extract code and test code from the API response.
- The system can compile the extracted code and test code.
- The system can extract compile errors.
- The system can execute the extracted code and test code.
- The system can extract the execution results.
- The system can extract execution errors.
- The system can combine the extracted errors and the generated code to create a prompt as a correction request and make a request to the ChatGPT API again.
- The system can extract the success of the tests.
- The system can present successful code and test results to users.

### 3.3.2. Sub Functions
- The system can save the API requests, responses, generated code, test results, and execution times to a database.
- Allow users to view past results.

## 4. System Development
### 4.1. Development Environment

The backend programming language is Java, the frontend is developed using HTML, CSS, and JavaScript, the database is MySQL, and the build tool is Gradle. The external API used is the ChatGPT API, and the development environment is VSCode. Testing is conducted using JUnit, and the web technologies used are JSP/Servlets with an MVC model structure.

### 4.2. System Implementation

The system's functionalities are divided into two main parts: code generation and result verification. Key classes include ProcessPromptServlet for handling code generation requests, ChatGPTAPISupport for API request processing, ResponseExtractor for processing API responses, and CodeExecutorSupport for code execution and evaluation.

### 4.2.1. Prompt Input Function

In this system, "user input" and "prompts originally set in the system" are sent to the API together. There are no restrictions on the language of the input prompt, but English is preferable to conserve the number of tokens in the input prompt. Tokens are counted as one token for each English word, but when input in Japanese, one character is treated as one token, and if kanji characters are used, several tokens are used for one character. When the user inputs the same content in Japanese and English, such as "Create a text-based Tetris" and "Please create a text-based Tetris," and makes an API request, the prompt_tokens keys in the respective API responses are "190" and "200." As can be seen from these results, using English prompts leads to reduction in the number of tokens, and there are benefits in terms of output prompts until the token limit is reached, as well as financial benefits, so prompts in this system are in English as much as possible.

### 4.2.2. API Request Function

A request is made by combining user input with a prompt that has been preset in the system.
In order to read the generated code from the API response, the following points should be reflected and must be read as part of the API request.
- Code and test code are generated.
- Code is written in the Java language.
- Code is generated between the [CODE] tag and the [/CODE] tag.
- Test code is generated between the [TEST] tag and the [/TEST] tag.

### 4.2.3. API Response Processing Function

JSON format responses undergo double escaping, so we made modifications to ensure proper escaping without impairing the readability of the code.

### 4.2.4. Code Extraction Function

The method for extracting code from the API response is as follows. Assuming that the content key of the API response contains the [CODE] tag, [/CODE] tag, [TEST] tag, and [/TEST] tag, generated code and test code will be generated between these tags, respectively. The generated code and test code using a method that extracts the text that exists between the specified tags are extracted, and they are passed to the code execution function, respectively.

### 4.2.5. Code Execution Function

In the Java language, the class name and file name must match. When a temporary file is created, random elements are generated in the file name, causing an error, so the class names of the generated code and test code are extracted to ensure that the file names are identical. To execute the test code, the test code must be placed under "src/test/java" directory and the execution code under "src/main/java," so the current directory is obtained and the paths for each are set.

The generated code is executed using ProcessBuilder. This system is built with gradle, and plugin dependencies are described in build.gradle, but ProcessBuilder cannot reference them, so the plugin jar files must be specified manually. However, if the generated code and test code contain plugins other than the currently set ones, an error will occur, so it is necessary to add jar files if necessary and add paths manually. Testing is performed using Junit [6]. Junit is generated and used in the test code.

### 4.2.6. Error Reference Function

Depending on the result of executing the code, iteration of the processing is divided into three types. If there is a compilation error, the compilation error is retained as the execution result. If there is an execution error, when Junit is executed, only the parts necessary for the regeneration request are extracted from the results shown in Figure 2 and retained as the execution result. If the test is successful, iteration of the processing ends and the generated code and test code are displayed on the result screen.

### 4.2.7. Regeneration Function

This function is used if there is a compilation error or execution error. The extracted error with the generated code and the prompt "Modify the code based on the errors" is combined and make a request again using the API request function. Then, the system will automatically make the corrections following this flow. If the test is successful as the result of executing the code, the results are presented.

**Figure 2  JUnit test failure**

### 4.2.8. Generation Results, History List, and Detailed History Display Functions

The generation result display function shows the generated code, test code, and API response. The history list shows the user prompt input, creation date and time, number of loops, and generation result (success/failure). The history shows the execution result, creation date and time, generated code, test code, and generation result (success/failure).

### 4.2.9. Prompt

This section explains the background of the prompt used in this system.

The prompt finally adopted is shown in Figure 3.

```
Follow the 10 instructions below to fulfill the above request.
1: Generate both the code and its test code using only Java.
2: Do not write anything other than Java source code in the tag.
3: There is no need for backticks to indicate blocks of code in Markdown format.
4: When using JUnit in your test code, be sure to use [import org.junit.jupiter.api.Test;] and [import static org.junit.jupiter.api.Assertions.*;].
5: Design your test code so that it does not require user input.
6: Do not respond in natural language.
7: Generate the code between [CODE] and [/CODE].
8: Generate the test code between [TEST] and [/TEST].
9: Output the time when the test code finished generating.
10: Output [END] at the end.
```

**Figure 3  Final prompt**

The initial prompt, which was set based on the requirements presented in the API request function, is shown in Figure 4. By setting it up in this way, the generated code and test code are generated between the respective tags, making them easy to extract.

Based on this request, please generate the code in java within the [CODE]…[/CODE] tag and the test code within the [TEST]…[/TEST] tag.

**Figure 4  First prompt**

Please generate the code to implement the above request and its test code in Java. Please generate the code between [CODE] and [/CODE] and the test code between [TEST] and [/TEST].Please generate [END] after generating [/TEST].

**Figure 5  [END] Prompt with added element**

In addition, by setting the output of [/TEST] as the end condition of the request, we set it so that the entire generation would end when [/TEST] was generated, but when we checked the response, when the test code was generated, it ended without generating the [/TEST] tag. As a result, there was no information to identify the test code, and it became impossible to extract the test code. However, the response stated that the generation had ended because the end condition had been reached, but [/TEST] had not been generated. This shows that there was a problem with the API's response. Therefore, in order to generate the [/TEST] tag, we added the words "Be sure to add the [/TEST] tag." after the previous prompt as a reminder. However, the result did not change, and the tag remained ungenerated.

Next, considering the possibility that the end condition would not be generated, we set the end condition to output

[END] and added the words "Please generate [END] after generating [/TEST]." to the prompt as shown in Figure 5. However, the result did not change, and neither [END] nor [/TEST] was generated. However, when we checked the response, it said that the end condition had been reached. From this result, it was thought that the system may not understand that tags such as [END] are generated (even though other tags are understood). Therefore, we thought that if the information added to the end of the response is not a tag, but is added in a completely different direction, the information will be ignored and the tag will be generated. Therefore, we added the phrase "Please output the time when the last generation finished." to the prompt, as shown in Figure 6. As a result, we succeeded in generating the [/TEST] tag. However, the response corresponding to this prompt was an inappropriate reply, "The generation finished at: [TIME]." Furthermore, even after adding this phrase, we often confirmed responses in which the tag itself was not generated correctly. A common feature in these cases was that natural language was generated at the beginning of the "content" of the response, such as "The code to implement the calculator program is as follows:" or "Sure! Here's an implementation of a basic calculator program in Java." Therefore, we thought that if we added a phrase to the prompt that instructs not to generate natural language, the tag would be generated correctly. Therefore, we added the phrase "Do not respond in natural language." As a result, the amount of natural language generated was dramatically reduced and the success rate of tag generation increased.

Based on this request, please generate the code in java within the [CODE]...[/CODE] tag and the test code within the [TEST]...[/TEST] tag. "Please output the time when the last generation finished."

**Figure 6    Prompt with time output added**

## 5.   Running the System

Figure 7 shows the detailed execution result when a user enters the prompt, "Weather Forecast App: A simple command line based weather forecast app. When a user enters a city name, it displays the current weather information for that city. Make an HTTP request to an external API (e.g. OpenWeatherMap), parse and display the JSON response." The loop is repeated twice.

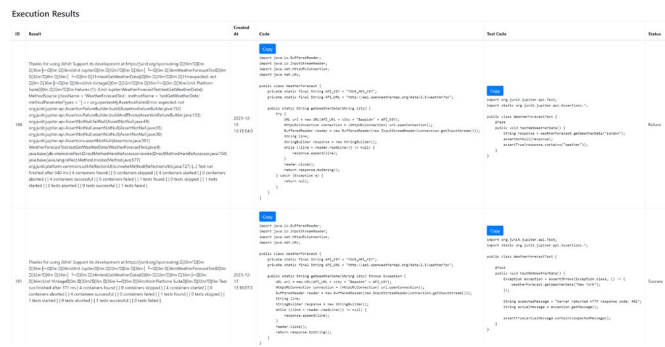

**Figure 7    The detailed execution result**

The failure of the first test means that the "getWeatherData" method returned null, suggesting that the response from the API failed or the data could not be retrieved for some reason. In response, the method was modified to throw an exception so that if there is a problem with the response from the API, the caller will be aware of the problem and can respond appropriately. The second test also changed the testing approach, and the new test expects the "getWeatherData" method to throw an exception under certain circumstances. This is to check whether certain error situations, such as invalid requests to the API or authentication errors, are handled properly. The revised code showed improved error handling, test accuracy, and practicality. However, the test code focused on exception handling and omitted test items for normal operation.

## 6.   Discussions

The system developed in this study is considered to meet the research goals. ChatGPT API requests often receive insufficient API responses, and ChatGPT is not omnipotent; it can do basic things, but it has a tendency to repeat small mistakes. It is a difficult question to determine to what extent these mistakes can be tolerated in the system, and there is a concern that it will rely on whimsical and random answers.

If the generated code is implemented using an external API, ChatGPT cannot obtain the API key on its own, so it is difficult to generate code with sufficient content. In such cases, the goal is to pass the test, and only tests for exception handling that can be implemented are generated, and important tests for success are insufficient. In this way, it is possible that the test content may be rough. In this system, we focused particularly on whether the code can be compiled and executed. However, we did not consider whether the test code was comprehensive. Therefore, it is considered necessary to incorporate the perspective of "test comprehensiveness" into the repeated generation, execution, and correction of this system. By improving the comprehensiveness of the tests, the possibility of successful code generation will increase, and it will be possible to promote the improvement of the quality of the generated code.

This system has an issue where if the generated code uses a plugin that is not covered by the system, an execution error will occur when it is compiled and executed. To eliminate this error, it is necessary to either configure a wide range of plugins in the system beforehand, or to have a function that automatically places the jar file of an unconfigured plugin in the system when it is detected. In addition, the current system is limited to testing source code written in Java with Junit and generating a single file that contains functional code such as string manipulation, mathematical calculations, and data structure manipulation. However, by using this system in parallel and implementing a system that combines the governing system with the entire governed code, tests the consistency between files, and corrects each file if it is inappropriate, it may be possible to develop a system that automatically generates web applications consisting of multiple files.

## 7. Conclusion

This study has developed an automatic correction system for source code generated using ChatGPT to automate and streamline the coding process. This system has the function of automatically executing the code generated by ChatGPT together with the test code, detecting and correcting compilation and execution errors. As a result, the user can obtain the code in the appropriate state by entering any prompt. The system developed in this study contributed to improving the quality of the code generated by ChatGPT and was shown to be an effective tool for improving the efficiency of programming work. However, fundamental limitations such as the data dependency and risk of generating false information of ChatGPT itself were also revealed. In addition, in order to test the generated code, the plugins used in the generated code must be set in the system, and it is necessary to consider how to deal with this problem.

Future challenges include overcoming these limitations and realizing a more advanced automatic correction function. In addition, it is believed that more accurate corrections can be made by focusing on the comprehensiveness of the test code as well as the generated code. Furthermore, in order to deal with errors in which tests fail when code is generated using plugins that are not expected by the system, it is considered necessary to prepare a wide range of plugins in advance on the system side and to have the function to add plugin jar files to the system at any time.

### Acknowledgements

This research was partly supported by Grant-in-Aid for Scientific Research (C) 21K1279.